\documentclass[aps,english,prb,reprint,superscriptaddress, citeautoscript,cite]{revtex4-1}

\usepackage{graphicx}
\usepackage{amssymb}
\usepackage{babel}
\usepackage{color}
\usepackage{multirow}

\begin{document}

\title{Exploring the Electronic and Magnetic Properties of New Metal Halides from Bulk to Two-Dimensional Monolayer: RuX$_3$ (X=Br, I)
\footnote{\dag~Electronic Supplementary Information (ESI) available: [See supplementary material for the exchange interaction parameters such as $J_1$, $J_2$ and $J_3$, phonon and electronic band structures, relative energy differences of bulk RuX$_3$, electronic density of states of monolayer RuX$_3$, relative energy differences for each configurations with respect to U and U+SOC parameters and compared band structures of monolayer RuX$_3$ in NM, FM, Neel, Stripy and Zigzag magnetic order using U+SOC methods.]}}

\author{F. Ersan}
\affiliation{Department of Physics, Adnan Menderes University, Aydin 09010, Turkey}
\affiliation{Department of Physics, Bilkent University, Ankara 06800, Turkey.}
\author{E. Vatansever}
\affiliation{Dokuz Eyl\"{u}l University, Faculty of Science, Physics Department, T{\i}naztepe Campus, 35390 İzmir, Turkey}
\author{S. Sarikurt}
\affiliation{Dokuz Eyl\"{u}l University, Faculty of Science, Physics Department, T{\i}naztepe Campus, 35390 İzmir, Turkey}
\author{Y.Y\"{u}ksel}
\affiliation{Dokuz Eyl\"{u}l University, Faculty of Science, Physics Department, T{\i}naztepe Campus, 35390 İzmir, Turkey}
\author{Y. Kadioglu}
\affiliation{Department of Physics, Adnan Menderes University, Aydin 09010, Turkey}
\affiliation{Department of Physics, Bilkent University, Ankara 06800, Turkey.}
\author{H. D. Ozaydin}
\affiliation{Department of Physics, Adnan Menderes University, Aydin 09010, Turkey}
\author{O.\"{U}zengi Akt\"{u}rk }
\affiliation{Department of Electrical and Electronic Engineering, Adnan Menderes University,
	09100 Ayd{\i}n, Turkey}
\affiliation{Nanotechnology Application and Research Center, Adnan Menderes University, Aydin 09010, Turkey}
\author{U. Ak{\i}nc{\i}}\email{umit.akinci@deu.edu.tr}
\affiliation{Dokuz Eyl\"{u}l University, Faculty of Science, Physics Department, T{\i}naztepe Campus, 35390 İzmir, Turkey}
\author{E. Akt\"{u}rk}\email{ethem.akturk@adu.edu.tr}
\affiliation{Department of Physics, Adnan Menderes University, Aydin 09010, Turkey}
\affiliation{Nanotechnology Application and Research Center, Adnan Menderes University, Aydin 09010, Turkey}
\date{\today}



\begin{abstract}
Theoretical and experimental studies present that metal halogens in MX$_3$ forms can show very interesting electronic and magnetic properties in their bulk and monolayer phases. Many MX$_3$ materials have layered structures in their bulk phases, while RuBr$_3$ and RuI$_3$ have one-dimensional chains in plane. In this paper, we show that these metal halogens can also form two-dimensional layered structures in the bulk phase similar to other metal halogens, and cleavage energy values confirm that the monolayers of RuX$_3$ can be possible to be synthesised. We also find that monolayers of RuX$_3$ prefer ferromagnetic spin orientation in the plane for Ru atoms. Their ferromagnetic ground state, however, changes to antiferromagnetic zigzag state after U is included. Calculations using PBE+U with SOC predict indirect band gap of  0.70 eV and 0.32 eV  for the optimized structure of RuBr$_3$ and RuI$_3$, respectively.  Calculation based on the Monte Carlo simulations reveal interesting magnetic properties of  $\mathrm{RuBr_{3}}$, such as large Curie temperature against $\mathrm{RuI_{3}}$, both in bulk and monolayer cases. Moreover, as a result of varying exchange couplings between neighboring magnetic moments, magnetic properties of $\mathrm{RuBr_{3}}$ and $\mathrm{RuI_{3}}$ can undergo drastic changes from bulk to monolayer. We hope our findings can be useful to attempt to fabricate the bulk and monolayer of RuBr$_3$ and RuI$_3$. 
\end{abstract}



\maketitle
\section{Introduction}
Recently, the family of transition metal trihalides MX$_3$, where M is a metal cation (M= Ti, V, Cr, Fe, Mo, Ru, Rh, Ir) and X is halogen anion (X= Cl, Br, I), have received increasing attention due to 
their potential applications in spintronics~\cite{angelkort2011temperature,wang2011electronic,wang2016doping,hillebrecht2004trihalides,zhou2016evidencing,li2016first,banerjee2016proximate}. Even though these materials have been known for  more than 50 years~\cite{tsubokawa1960magnetic,fletcher1963anhydrous,baker1964magnetic,lin1993dimensional}, 
and their structure is well-investigated; only a few 
three-dimensional (3D) layered transition metal halides have been observed experimentally\cite{troyanov1991x,miro2014atlas}. 
In recent years, it is possible to exfoliate these 3D layered crystals 
down to two-dimensional (2D) monolayers, due to the weak interlayer van der Waals (vdW) interactions~\cite{bengel1995tip,plumb2014alpha}. 
For instance, Weber \textit{et al.}~\cite{weber2016magnetic} report
the exfolation of the magnetic semiconductor $\alpha$-RuCl$_3$ into the first halide monolayers and investigations of its in-plane structure show that it is retained during the exfolation 
process. Huang \textit{et al.} use magneto-optical Kerr effect (MOKE) microscopy to demonstrate the monolayer CrI$_3$ is an Ising model ferromagnet with out-of-plane spin orientation. They 
find out that its Curie temperature of 45 K is only slight lower than 61 K of the bulk crystal~\cite{dillon1965magnetization}, consistent with a weak interlayer coupling~\cite{huang2017layer}. 
Similarly very recently, McGuire \textit{et. al}\cite{mcguire2017high} both 
experimentally and theoretically focus 
on the crystallographic and magnetic properties of transition metal compound $\alpha$-MoCl$_3$ behaivor above the room temperature. 

Transition metal trihalides provide a rich family of materials with a wide range of electronic, optical and mechanical properties in which also low dimensional magnetism 
can be examined, and therefore rapidly increasing theoretical
researches exists on this area\cite{zhang2015robust,he2016unusual,kuzubov2017dft,mcguire2017crystal,iyikanat2018tuning}. 
In our previous study, we systematically investigate the electronic and magnetic properties of an $\alpha$-RuCl$_3$ monolayer 
using DFT and MC simulations\cite{sarikurt2018electronic}, and our cleavage energy calculations give smaller value than that of graphite, which means that the $\alpha$-RuCl$_3$ monolayer 
can be easily obtained from its bulk phase and also we find that it is stable 2D intrinsic ferromagnetic semiconductor. Similarly, a class of 2D ferromagnetic monolayers CrX$_3$ (X= Cl, Br, I) 
is studied by Liu \textit{et al.}\cite{liu2016exfoliating} by using first principle calculations combined with MC simulations based on the Ising Model. 
They confirm that the feasibility of exfoliation from 
their layered bulk phase by the small cleavage energy and all the ground states monolayers are semiconducting with band gaps of $2.28$, $1.76$ and $1.09$ eV for CrCl$_3$, CrBr$_3$, CrI$_3$, 
respectively. Furthermore, the estimated Curie temperatures for CrCl$_3$, CrBr$_3$, CrI$_3$ are found $66$, $86$, $107$ K, respectively. Different from this study, among 
the chronium trihalides, the CrI$_3$ is also studied by another group both in experimentally and theoretically\cite{mcguire2015coupling} since it is the simplest to prepare 
due to iodine can be handled relatively easy solid at room temperature. They find that an easily cleavable, layered and insulating ferromagnet with Curie temperature of 61 K. Similarly, 
Huang \textit{et al.}~\cite{huang2017quantum} examine RuX$_3$ (X=Cl, Br, I) monolayers and use only RuI$_3$ monolayer as an exemplary material to study their electronic and magnetic 
properties by using first-principle calculations. Their result reveal that the ground state of the RuI$_3$ monolayer is ferromagnetic with estimated Curie temperature to above the room 
temperature $\sim$360 K. Nevertheless, \textit{ab-initio} molecular dynamics (AIMD) simulations confirm its thermal stability at 500 K and also a clear Dirac cone in the spin-down channel 
appears at the $K$-point in the Brillouin zone near the Fermi level of its band structure. Similarly, relying on our previous experience\cite{sarikurt2018electronic}, in this work we both focus on 
from bulk to monolayer RuI$_3$ and RuBr$_3$ electronic and magnetic properties in detail. Our results which are systematically investigated below are incompatible with the previous study. 
Since, our theoretical results demonstrate that RuBr$_3$ and RuI$_3$ can be stable in bulk form and monolayers of them can be obtained from their bulk phases by cleavage methods. 
We have obtained the possible magnetic ground state for bulk and monolayer forms of RuBr$_3$ and RuI$_3$ using PBE, PBE+SOC and U+SOC calculations. 
The FM spin orientation is the most favorable configurationfor PBE and PBE+SOC results,  while Zigzag spin orientation is favored after adding the Hubbard parameter. Hence, with considering the Hubbard U correction, the favored spin orientation can be altered to  antiferromagnetic zigzag situation and these RuX$_3$ monolayer structures have suitable band gaps for various  optoelectronic device applications.
Afterwards, we have obtained the magnetic exchange coupling constants and magnetic anisotropy energies from  the density functional theory calculations. Using these parameters, we have also performed Monte Carlo (MC) simulations, and estimated the Curie temperatures for $\mathrm{RuBr_{3}}$ and $\mathrm{RuI_{3}}$. According to our MC data, both structures in bulk and monolayer forms are found to be magnetically ordered at temperatures well below the room temperature.

\section{Computational Methodology}\label{comp}
Density functional theory (DFT) calculations were performed by using VASP package~\cite{vasp1,vasp2} within generalized gradient approximation (GGA)~\cite{Perdew1992atoms}. The Perdew-Burke-Ernzerhof (PBE) functionals were used for the exchange-correlation potential~\cite{pbe} and the Projector Augmented Wave (PAW) pseudo potentials were adopted~\cite{paw1,paw2}. A cutoff energy of 400 eV for the plane wave basis set was used. Monkhorst-Pack~\cite{monk} mesh of 16$\times$8$\times$1 (for bulk) and 16$\times$8$\times$1 (for monolayer) were employed for the Brillouin zone integration. A supercell with a 24 {\AA} vacuum distance was used in order to avoid interactions between two adjacent monolayer system when the periodic conditions are employed. The geometrical configurations were optimized by fully relaxing the atomic structures, until Hellmann-Feynman forces acting on each atom is reduced to less than 0.002 eV/{\AA}. The convergence of the total energy is achieved until the energy difference between successive iteration steps are less than $10^{-5}$ eV. Phonon dispersion curves were obtained by PHONOPY code \cite{phonopy} for the 2$\times$2$\times$1 supercell and displacement of 0.01~{\AA} from the equilibrium atomic positions. Finite temperature AIMD calculations within Verlet algorithm were performed for thermal stability test. We used Nos\'{e} thermostat for the duration of 2 picoseconds (ps) at 500 K for 3$\times$3$\times$1 $\alpha$-RuX$_3$ (X=Br, I) supercells. 

To elucidate the magnetic structure of RuX$_3$, and the nearest, second-nearest, and third-nearest neighbor exchange-coupling parameters (J$_1$, J$_2$ and J$_3$, respectively), we adapt the total energy values obtained from DFT calculations for different magnetic configurations to the Heisenberg Spin Hamiltonian:
\begin{eqnarray}\label{mc_eq1}
\nonumber
\mathcal{H}&=&-J_{1}\sum_{<ij>}\mathbf{S}_{i}.\mathbf{S}_{j}-J_{2}\sum_{<<ik>>}\mathbf{S}_{i}.\mathbf{S}_{k}-J_{3}\sum_{<<<il>>>}\mathbf{S}_{i}.\mathbf{S}_{l}\\
&&+k_{x}\sum_{i}(S_{i}^{x})^{2}+k_{y}\sum_{i}(S_{i}^{y})^{2}, 
\end{eqnarray}
where $\mathbf{S}_i$ is the spin at the Ru site i and (i, j), (i, k) and (i, l) stand for the nearest, second-nearest, and third-nearest Ru atoms, respectively. And $k_{x}$ and $k_{y}$ denote the out-of-plane magnetic anisotropy constants, respectively. The numerical values of $k_{x}$ and $k_{y}$ are obtained from magneto-crystalline anisotropy energies (MAE).


By mapping the DFT energies to the Heisenberg Hamiltonian, J$_1$, J$_2$ and J$_3$ can be calculated from following equations~\cite{Sivadas2015}:
\begin{equation}
E_{FM/Neel}=E_0 - (\pm 3J_1 + 6J_2 \pm 3J_3)S^2
\end{equation}
and
\begin{equation}
E_{Zigzag/Stripy}=E_0 - (\pm J_1 - 2J_2 \mp 3J_3)S^2
\end{equation}

The calculated J$_1$, J$_2$, J$_3$ exchange-coupling parameters, in-plane (E[100]-E[010]) and out-of-plane (E[100]-E[001]) MAEs and magnetic anisotropy constants can be found in Supplementary Material (S.M) Table S3 and S4 for both RuBr$_3$ and RuI$_3$. The Curie temperature was calculated by using these exchange-coupling parameters in MC simulations based on the Heisenberg model.

\section{From Bulk to Two-Dimensional Monolayer RuX$_3$; DFT calculations} 
\begin{figure*}[!htbp]
	\includegraphics[scale=0.92]{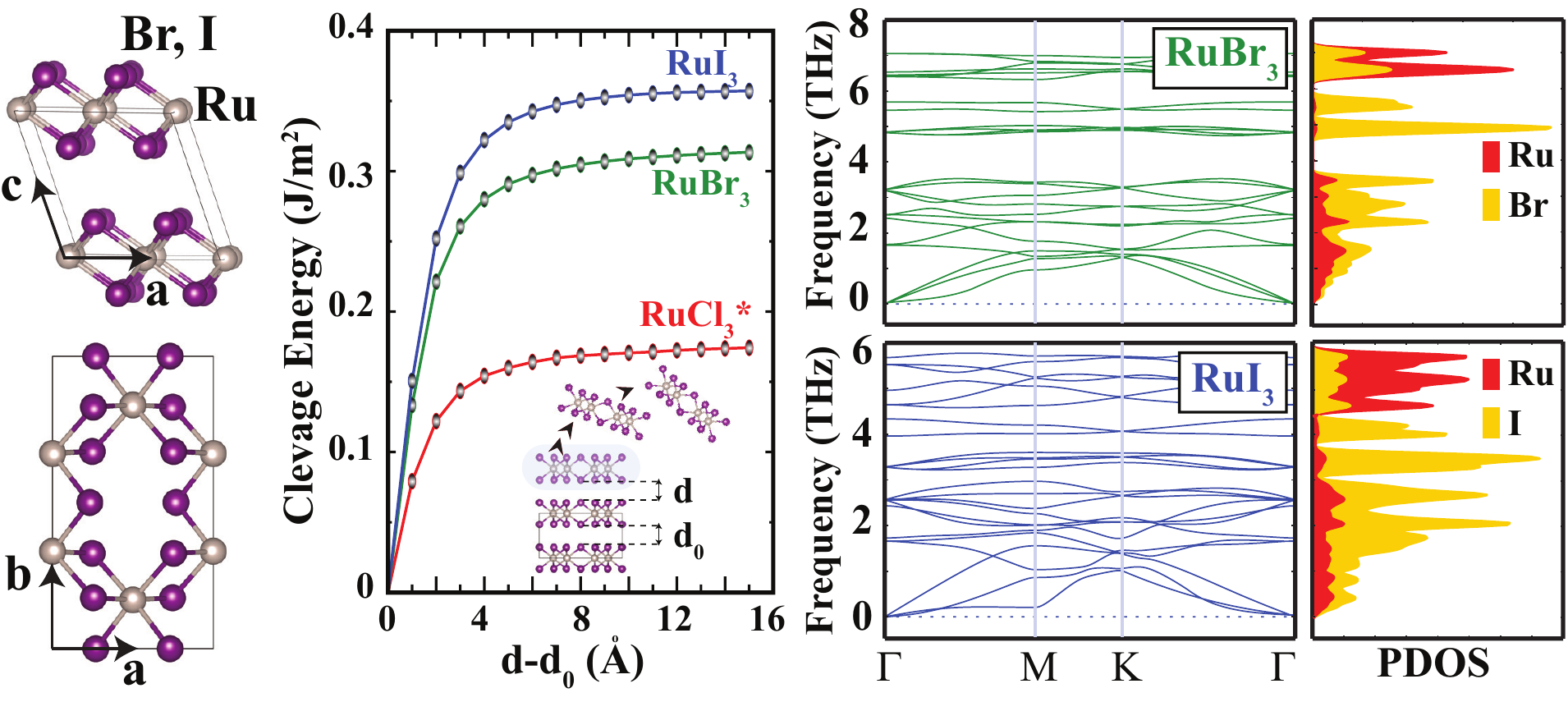}
	\caption{(Color online) Top and side view of bulk (P3$_1$12 space group) RuX$_3$ structure is shown at left panel. Middle panel for cleavage energy as a function of separation between the two fractured parts. The fracture distance is denoted as d and the equilibrium interlayer distance of ruthenium trihalides as d$_0$. 
		Inside the graph: side view of bulk a-RuX$_3$ used to simulate the exfoliation procedure, RuCl$_3$ results taken from Sarikurt \textit{et al.} study~\cite{sarikurt2018electronic}. Phonon band structures and corresponding PDOS of hexagonal RuX$_3$ monolayer structures are illustrated at right panel.}
	\label{cleavage}
\end{figure*}

\begin{figure*}[!htbp]
	\includegraphics[scale=0.6]{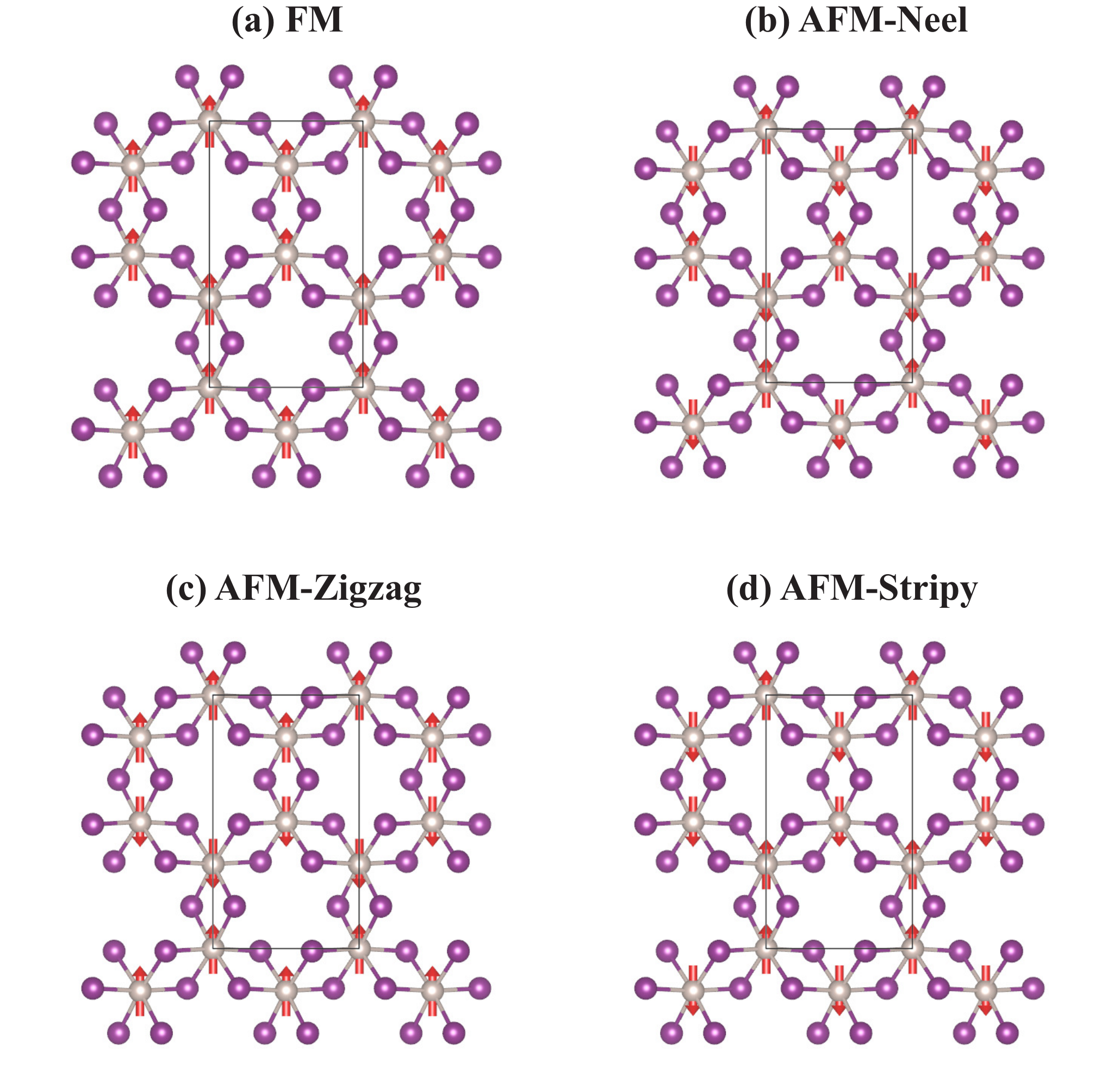}
	\caption{(Color online) Different spin configurations of the RuX$_3$ structures: (a) FM ordered, (b) AFM-Ne\'{e}l ordered, (c) AFM-Zigzag ordered and (d) AFM-
		stripy ordered.}
	\label{magneticorder}
\end{figure*}

Transition metal halides can be observed in several types of space groups such as C2/m, Pmnm, P6$_3$/mcm, P3$_1$12. Among them metal halide crystal structure 
in P3$_1$12 space group has equidistant metal atoms in the cell. Experimentally RuBr$_3$ can have Pmnm space group at low temperature while it has 
P6$_3$/mcm space group at high temperatures, and RuI$_3$ has P6$_3$/mcm space group at room temperature.\cite{Angelkort2009} In this paper we study only 
bulk RuBr$_3$ and RuI$_3$ structure in P3$_1$12 space group which is valid for RuCl$_3$ (see Fig.~\ref{cleavage}). And also we obtain and investigate their stable monolayer 
forms. We initialy constructed the bulk RuBr$_3$ and RuI$_3$ structures, and we obtained the optimized lattice constans as a=6.25 {\AA}, b=10.83 {\AA},
c=6.31 {\AA} for RuBr$_3$, a=6.77 {\AA}, b=11.67 {\AA} and c=6.71 {\AA} for RuI$_3$. Since they have not been sythesized in RuCl$_3$ bulk type, we expose them in 
dynamical stability tests such as phonon and molecular dynamic (MD) calculations. Obtained phonon band structures and corresponding thermodynamic variables 
are given in S.M Fig.S1. As can be seen in Fig.S1 both of bulk RuX$_3$ structures are dynamically stable for P3$_1$12 space group, and 
their heat capacities trend follow the Dulong-Petit limit after around 200 K. The AIMD calculations also showed that bulk form of RuX$_3$ 
structures are thermally stable at 500K for 2 ps. After optimization and stability calculations we examine their electronic properties, according to 
standard PBE calculations we found that both of bulk RuX$_3$ structures are metal. Bader charge analysis indicates that each Ru atom in the bulk RuBr$_3$ gives 
0.70 electrons (e$^-$) and each Br atom takes 0.23 e$^-$. These values are 0.30 e$^-$ for Ru atoms and 0.10 e$^-$ for I atoms in the bulk RuI$_3$ structure.
To examinate the favorable spin oriented status in the bulk RuX$_3$ structures four types of spin configurations are considered (FM, AFM-Ne\'{e}l, AFM-Zigzag
and AFM-Stripy) for ruthenium atoms as seen in Fig.~\ref{magneticorder}. We performed these calculations for three different DFT methods such as PBE, PBE+SOC 
(spin-orbit coupling) and U+SOC (for Hubbard U=1.5 eV) calculations. According to the calculations FM spin orientation is favorable for PBE and PBE+SOC results, 
while Zigzag spin orientation is favored after adding the Hubbard parameter (please see S.M Table S1 for relative energies, and band structures of bulk RuX$_3$).
Finally, we tested the possibility of the 
exfoliation techniques to get few layers of monolayer from their bulk forms. For these calculations bulk RuX$_3$ structures are extended in z-direction 
and four layered RuX$_3$ structures are created, and then we implemented a fracture in the bulk after four periodic layers and systematically increased 
this fracture distance; at the end we calculated the corresponding cleavage energy (CE) (Fig.~\ref{cleavage}). RuCl$_3$ results are taken from our 
previous study~\cite{sarikurt2018electronic}. As can be seen increasing of the halogens rows in the periodic table enhances the cleavage energy. 
But calculated energies are comparable with graphite, and other MX$_3$ materials~\cite{Tan_2017,liu2016exfoliating,tsubokawa1960magnetic,mcguire2015coupling,wang2016doping,zhang2015robust,he2016unusual}. 

\begin{figure*}[!htbp]
	\includegraphics[scale=0.6]{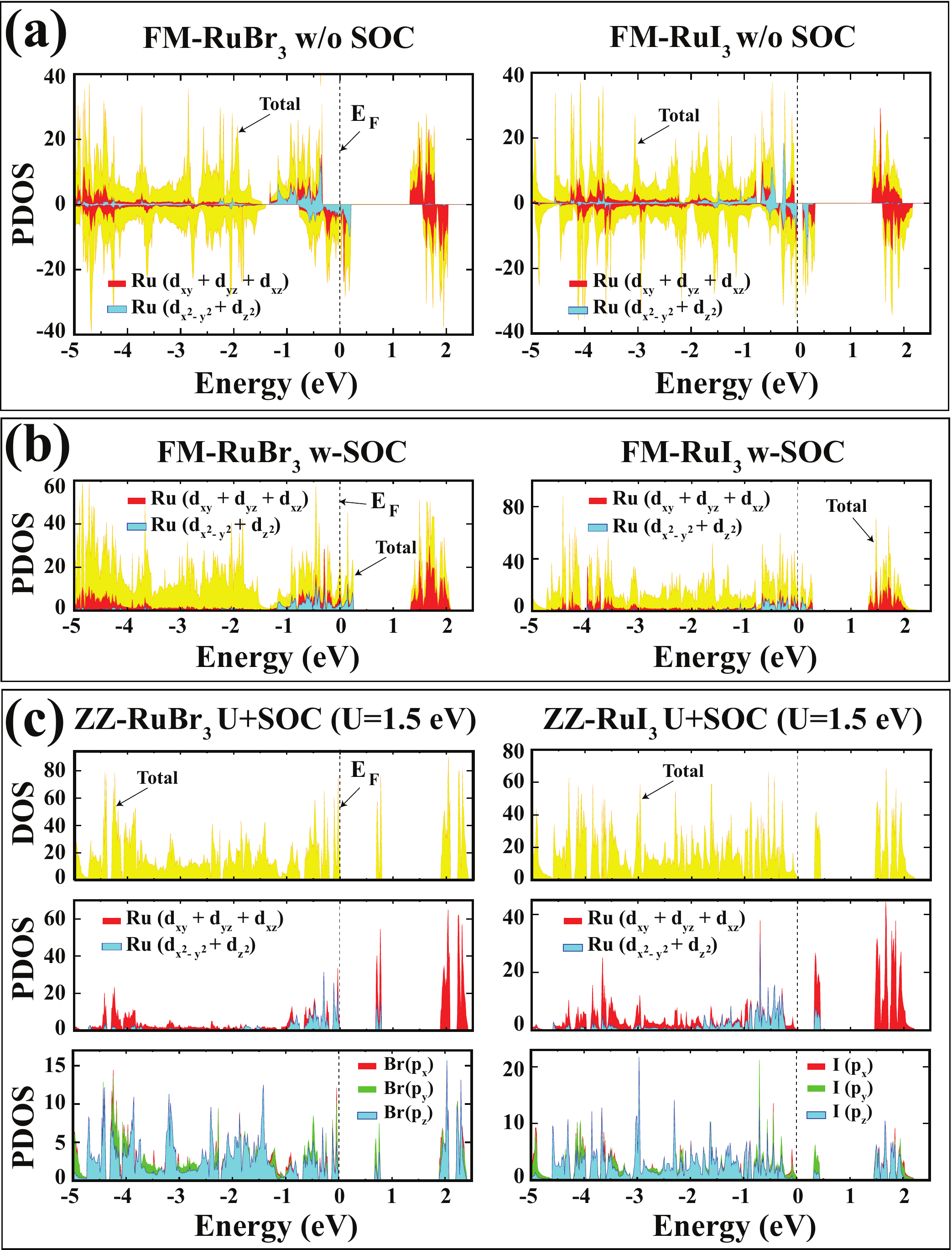}
	\caption{(Color online) Electronic PDOS (states/eV) of monolayer RuBr$_3$ and RuI$_3$ a) Total and $d$ orbital contribution in DOS for 
		FM-RuX$_3$ which calculated by PBE, b) otal and $d$ orbital contribution in DOS for 
		FM-RuX$_3$ which calculated by PBE+SOC, c) Total density of states, partial orbital contribution of ruthenium and halogen atoms are given 
		separately for ZZ-RuX$_3$ which is calculated with Hubbard U+SOC (U=1.5 eV) effect.}
	\label{phonon}
\end{figure*}

Monolayer RuBr$_3$ and RuI$_3$ structures are constructed in hexagonal unitcell, which have lattice constants of a=6.25 {\AA}, and a=6.78 {\AA} for 
RuBr$_3$ and RuI$_3$, respectively. This lattice value is 5.92 {\AA} for RuCl$_3$\cite{sarikurt2018electronic} as expected lattice constants increase 
by increasing the atomic radii from chlorine to iodine. Pauling electronegativity values are 2.20 for Ru atom 2.96 for Br and 2.66 for I atoms, this 
electronegativity difference results more electron transferring from Ru atoms to Br atoms than I atoms. According to Bader charge analysis while each Ru 
atom in RuBr$_3$ loses 0.72 electrons (e$^-$), this value is 0.32 e$^-$ in RuI$_3$ monolayer. This charge transfer interpretable such as there is 
more strength bond between Ru and Br atoms according to Ru-I atoms. Similar dynamical tests which are performed for bulk RuX$_3$ structures are also performed for
monolayers. Phonon band structures and their partial density of states (PDOS) of RuX$_3$ monolayers are illustrated in Fig.~\ref{cleavage}. Phonon dispersions are obtained by 
using PHONOPY code which is based on density functional perturbation theory as implemented in VASP. As can be seen, all phonon branches have positive frequency values in the whole Brillouin Zone (BZ) which implies the dynamical stability at T$\sim$0K. As mentioned later, spin-polarization is more effective in RuBr$_3$ 
with respect to RuI$_3$ monolayer. Thus, phonon band structure of RuBr$_3$ obtained with spin-polarized calculation due to it has large imaginary frequencies for 
out-of-plane acoustical branch (ZA) for spin-unpolarized status. In addition, phonon band structure of RuI$_3$ monolayer has a local minimum at the M high symmetry point for the ZA, which is associated with Kohn anomalies. Thermal stability tests are performed by AIMD calculations. All RuX$_3$ structures 
subjected to 500K temperature for 2 ps. At the end of calculations both of RuBr$_3$ and RuI$_3$ monolayers preserved their optimized 
atomic configuration which are obtained at T=0 K calculations. This means that RuX$_3$ monolayers can be stable at room temperature and
at least slightly above it. This conclusion is very important to utilize them in device technology. After the stability tests we start to investigate to 
determine their favorable magnetic ground states. For this examination, we changed the hexagonal RuX$_3$ unitcell to the rectangular cell and considered four types of spin configurations similar to bulk ones as seen in Fig.~\ref{magneticorder}. We performed geometric 
optimization calculations to the structures for all considered magnetic orientation status untill the pressure on the cell is approximately zero, with and without spin-orbit coupling (SOC) effect. According to these PBE and PBE+SOC calculations we found that FM state is energetically favorable spin oriented status for both RuBr$_3$ and 
RuI$_3$ monolayers. But AFM-Stripy-RuBr$_3$ has only 86 meV higher energy than FM state and this difference is reduced to 67 meV when the SOC is added 
in calculations. SOC is more effective in RuI$_3$ monolayer, energy difference between FM and Stripy state is 119 meV without SOC effects, while it 
becomes 10 meV with SOC contribution. Relative energy differences for other spin orientation states can be found in S.M Table S2. Each 
FM-RuX$_3$ structure has 4$\mu_B$ magnetic moment in per rectangular cell and each Ru atom in the cell has 1$\mu_B$ magnetic moment. We also calculated 
the cohesive energies of FM-RuX$_3$ structures to determine the strength of cohesion between the Ru and X atoms and we estimate 13.67 eV and 12.81 eV 
for per RuBr$_3$ and RuI$_3$ quartet atoms, respectively. Dominant orbital contribution to the electronic structure comes from Ru $d$ and halogen $p$ orbitals, 
Fig.~\ref{phonon} shows the electronic PDOS of RuX$_3$ monolayers for various spin-orientation and with (w) and without (w/o) SOC effect. 
As seen in Fig.~\ref{phonon} a) both of monolayers have large band gap for 
spin up channel, ordered in 1.65 eV and 1.45 eV for FM-RuBr$_3$ and FM-RuI$_3$, while density of states are very close to each other for spin down channels 
(There is a 20 meV gap between the DOSs for FM-RuBr$_3$, while this gap reaches to 100 meV for FM-RuI$_3$). 
For FM-RuBr$_3$ spin up state, two fold $e_g$ ($d_{z^2}$ and $d_{x^2}-d_{y^2}$) orbitals and three fold $t_{2g}$ ($d_{xy}$, $d_{yz}$ and $d_{xz}$) orbitals 
contribute equally to the valence band maximum (VBM), while there are just $t_{2g}$ orbitals contribution in conduction band minimum (CBM) and between 
1.7-1.9 eV. Also dominant contribution comes from $t_{2g}$ orbitals for spin down channel around the Fermi level, and again there are only $t_{2g}$ 
orbitals between 1.8-2.0 eV in spin down. Br atom $p$ orbitals give approximately equal contribution around VBM and CBM for both spin up and down channels 
(see S.M Fig.S3). For FM-RuI$_3$ spin up two fold $e_g$ orbitals are dominant at VBM and at CBM, $t_{2g}$ orbitals contributions start $\sim$0.2 eV 
lower energy from VBM, while there are not in CB. Spin down states posses similar situation with FM-RuBr$_3$ spin down channel. 
In plane $p$ orbitals ($p_x$, $p_y$) of iodine give major contribution to the VBM for spin up state as seen in Fig.S3. By including SOC effect in calculation 
for FM-RuBr$_3$ system gains metallic character, while FM-RuI$_3$ preserves semiconducting behavior (Fig.~\ref{phonon}b). Electric and magnetic properties 
of such layered metal halides must be investigated by including Hubbard \textit{U} correction term to the calculations, so we added \textit{U} from 0.5 
to 3.0 eV which increases by successive 0.5 eV value and we determined the favorable magnetic ground states for each added \textit{U} terms, we also 
repeated these calculations by adding \textit{U}+SOC terms in our calculations. Relative ground state energy graphs can be found in S.M. Fig.S4. 
According to our extended calculations, FM spin orientation is favorable just for U=0.5 eV both with and without SOC effect. For larger Hubbard energies 
zigzag (ZZ) orientation has minimum ground state energy comparing to others. We attained very close band gap value with experimentally obtained thin layered 
$\alpha$-RuCl$_3$ result~\cite{ziatdinov2016atomic} in our previous band structure calculations for RuCl$_3$ monolayer for Hubbard U=1.5 eV, 
thus we give in detail density of states for energetically favorable ZZ-RuX$_3$ (U+SOC and U=1.5 eV) monolayers in Fig.~\ref{phonon}c. As can be seen in Fig~\ref{phonon}c 
Hubbard U and SOC effects enhance the band gaps for RuX$_3$ monolayers and reaches 0.70 eV for RuBr$_3$ and 0.32 eV for RuI$_3$. While $t_{2g}$ orbitals 
of Ru atoms and $p_x$, $p_y$ orbitals of I atoms determine the VBM level, all orbitals of Ru and Br atoms approximately give similar contribution at VBM. At 
conduction band minimums $t_{2g}$ orbitals of Ru atoms are dominant. Calculated electronic band structures for all optimized RuX$_3$ monolayers, and also band trends can be found in S.M Fig. S5-S7.

\section{From Bulk to Two-Dimensional Monolayer RuX$_3$; Monte Carlo calculations} 
\subsection{Heisenberg-Kitaev Models}
Recently, magnetic properties of certain materials exhibiting strong SOC have been modeled by using the Heisenberg-Kitaev (HK) model \cite{kitaev}. For instance,
magnetic behaviors of $\mathrm{\alpha-RuCl_{3}}$ and $\mathrm{Na_{2}IrO_{3}}$ have been studied by Janssen et al.~\cite{janssen1}. They have demonstrated 
that the response of the system to an external field differs substantially for the different scenarios of stabilizing the zigzag state. The same group have also 
studied the honeycomb lattice HK model in an external magnetic field, and mapped out the classical phase diagram for different directions of the magnetic field~\cite{janssen2}. 
In addition, magnetic behavior and phase diagrams of iridium oxides $\mathrm{A_{2}IrO_{3}}$  have been investigated by Chaloupka and coworkers \cite{chaloupka1,chaloupka2} and by Singh et al.~\cite{singh}.
The latter group have demonstrated that the magnetic properties of $\mathrm{A_{2}IrO_{3}}$ can be modeled by using HK model including next-nearest neighbor interactions.

Apart from these works, there also exist several works dedicated to the investigation of magnetic properties of HK model in detail. For instance, 
the topological properties of the expanded classical HK model on a honeycomb lattice have been investigated
by Yao and Dong \cite{yao}. The effect of the spatially anisotropic exchange couplings on the order-disorder characteristics of HK model has been clarified by Sela et al. \cite{sela}.
Classical HK model on a  triangular lattice including the next-nearest neighbor interactions and  single ion anisotropy has been investigated by Yao \cite{yao2}.
Price and Perkins \cite{price} elaborated the finite temperature phase diagram and order-disorder transitions of classical HK model on a hexagonal lattice. In a separate work \cite{price2}, they have also 
studied the critical properties of the HK model on the honeycomb lattice at finite temperatures in which they have found that the model undergoes two phase transitions as a function of temperature. 
Finally, the relation between the classical HK model and quantum spin-$S$ Kitaev model for large $S$ has been discussed by Chandra et al. \cite{chandra}.

Although Ising model is often utilized in determination of magnetic properties of real magnetic materials \cite{huang2017quantum}, one may desire to take into account the apparent effect of MAE (see S.M Table S4)
in the atomistic spin model calculations. Hence, for simplicity, we base our simulations on an anisotropic Heisenberg model. Although, classical 
Heisenberg model is a simple model in comparison to HK model, it provides physically more reasonable
results in comparison to conventional Ising model which is only suitable for highly anisotropic magnetic systems.

\subsection{Monte Carlo Simulation Details}
In order to clarify the magnetic properties of $\mathrm{RuX_{3}(X=Br,I)}$, we proposed an atomistic spin model, and performed MC simulations based on the Metropolis algorithm \cite{newman} 
on a two dimensional honeycomb lattice with lateral dimensions $L_{x}=L_{y}=100$ which contains $N=10^{4}$ spins. We run our simulations based on the Hamiltonian defined by Eq.(\ref{mc_eq1}).
The numerical values of system parameters have been provided in S.M Table S3 and S4.
According to Eq. (\ref{mc_eq1}), a Ru atom with a pseudo spin $|\mathbf{S}_{i}|=1/2$ resides on each lattice site. We can briefly outline the simulation procedure based on Eq. (\ref{mc_eq1}) as follows:
Starting from a high temperature spin configuration, we progressively cool down the system until the temperature reaches $T=10^{-2}K$. We performed sequential spin flip update in our calculations with $10^{5}$ MC steps per site where
$10\%$ of this value have been discarded for thermalisation. Periodic boundary conditions (PBC)
were imposed in all directions. In order to reduce the statistical errors, we performed 100 independent runs at each temperature. Error bars were calculated using the Jackknife method \cite{newman}. During the simulation, the following 
physical properties have been monitored:
\begin{itemize}
 \item  Time series of the spatial components of total magnetization 
 \begin{equation}\label{mc_eq2}
 m_{\alpha}(t)=\frac{1}{N}\sum_{i=1}^{N}g\mu_{B}S_{i}^{\alpha}, \quad \alpha=x,y,z
 \end{equation}
 where $g$ is the Land\'{e} factor, and $\mu_{B}$ is the Bohr magneton. Using Eq. (\ref{mc_eq2}), we can obtain the thermal average of the magnitude 
 of the total magnetization vector $M_{T}$, as well as its components $M_{\alpha}$ according to the following relations 
 \begin{equation}\label{mc_eq3}
 \left\langle M_{\alpha}\right\rangle=\left\langle m_{\alpha}(t)\right\rangle, \quad \left\langle M_{T}\right\rangle=\left\langle \sqrt{\sum_{\alpha=x,y,z}m_{\alpha}^{2}(t)}\right\rangle. 
 \end{equation}
 \item In order to locate the transition temperature, we have also calculated the thermal average of magnetic susceptibility $\chi$ and magnetic specific heat as follows
\begin{equation}\label{mc_eq4}
\chi=N(\left\langle M_{T}^{2} \right\rangle-\left\langle M_{T}\right\rangle^{2})/k_{B}T,
\end{equation}
\begin{equation}\label{mc_eq5}
C=N(\left\langle \mathcal{H}^{2} \right\rangle-\left\langle \mathcal{H}\right\rangle^{2})/k_{B}T^{2}.
\end{equation}
where $k_{B}$ is the Boltzmann's constant. For the sake of completeness, we have also calculated the specific heat via 
\begin{equation}\label{mc_eq6}
C=\frac{\partial \left\langle \mathcal{H}\right\rangle}{\partial T}.   
\end{equation}
\end{itemize}

\subsection{Monte Carlo Simulation Results}
\begin{figure*}[!htbp]
	\includegraphics[scale=0.32]{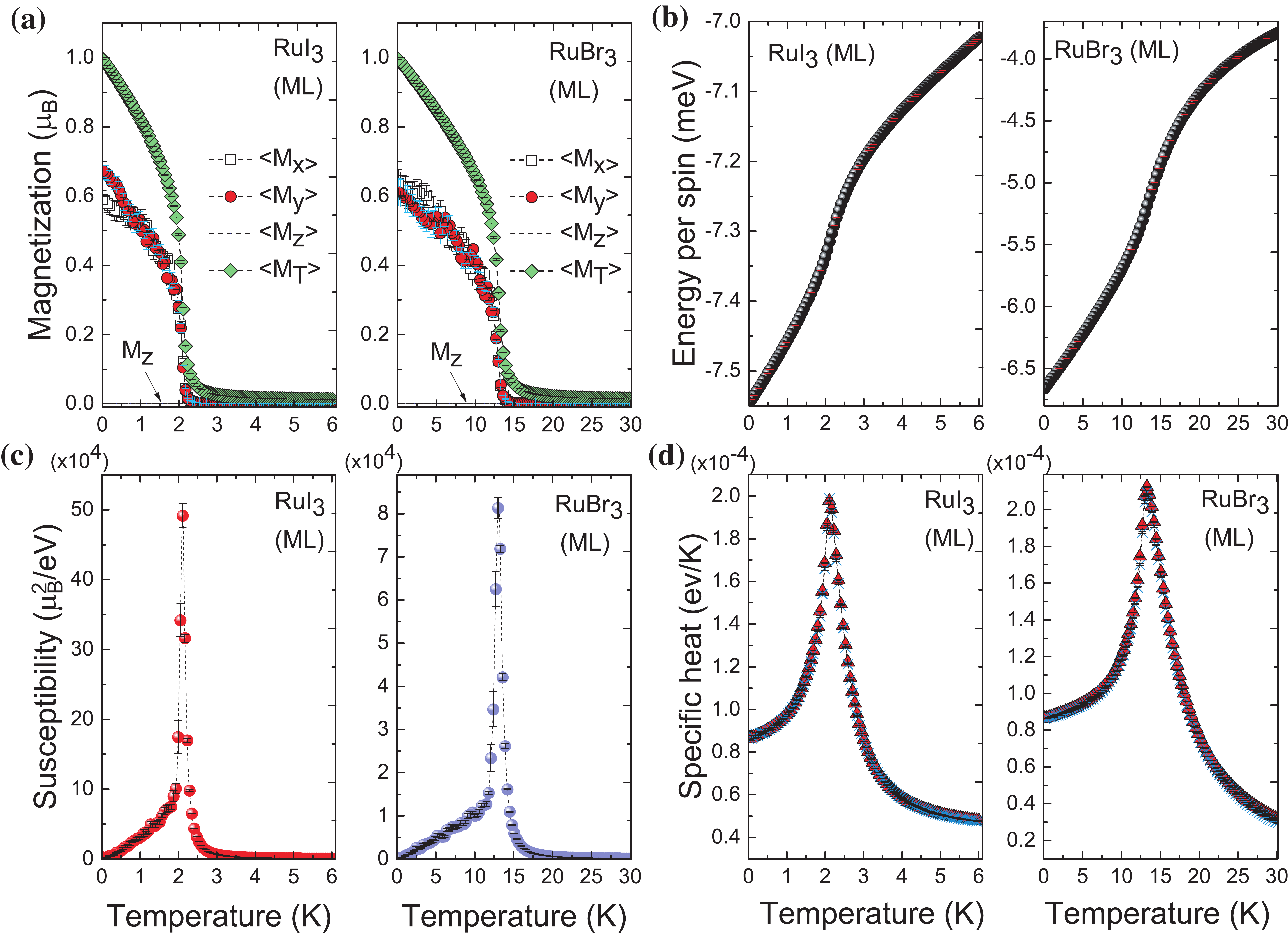}
	\caption{Temperature dependence of (a) average magnetization $M_{T}$ and its components $M_{\alpha}$, (b) average internal energy per spin,  (c) magnetic susceptibility 
		and (d) specific heat for $\mathrm{RuI_{3}}$ and $\mathrm{RuBr_{3}}$. In (d), different symbols denote the two distinct measurement methods for 
		specific heat as discussed in Eq. (\ref{mc_eq5}) $(\triangle)$ and Eq. (\ref{mc_eq6}) $(\times)$.} 
	\label{mc_fig} 
\end{figure*}

In Fig. \ref{mc_fig}, we display the MC simulation results regarding the magnetic properties of simulated $\mathrm{RuX_{3}}(\mathrm{X=Br,I})$ monolayer systems. In Fig. \ref{mc_fig}a, we plot the magnetization versus temperature for both structures.
As seen in this figure, starting from high temperature configuration, as the temperature gradually decreases then the non zero magnetization components emerge. Since the out-of-plane anisotropy constants $k_{x}$ and $k_{y}$ equal to each other,
main contribution to the total magnetization equally comes from $x$ and $y$ components whereas $z$ component does not contribute to the magnetic behavior. Although the components exhibit apparent fluctuations in the considered temperature range,
the magnitude $\left\langle M_{T}\right\rangle$ of total magnetization exhibits rather smooth behavior with error bars smaller than the data points. 
At very low temperatures $\left\langle M_{T}\right\rangle$ saturates to unity which means that $\mathrm{RuX_{3}}(\mathrm{X=Br,I})$ system exhibits ferromagnetic behavior at the ground state. 
This is consistent with results of our DFT calculations where we predicted that the stable ground state of  $\mathrm{RuX_{3}}(\mathrm{X=Br,I})$ is FM.
Thermal variation of internal energy is shown in Fig. \ref{mc_fig}b. 
Absolute value of $\left\langle \mathcal{H}\right\rangle$ at low temperature region is larger than that of the high temperature region. This is due to the fact that with increasing temperature, thermal fluctuations are enhanced, 
and the system evolves towards the paramagnetic regime. 
The transition temperature of $\mathrm{RuX_{3}}$ monolayers can be determined by examining the magnetic susceptibility and specific heat curves which are 
depicted in Figs.\ref{mc_fig}c and \ref{mc_fig}d. As seen in these figures, both response functions exhibit a peculiar peak in the vicinity of the magnetic phase transition temperature. According to our simulation results, transition
temperature values separating the ferromagnetic phase from paramagnetic phase are found to be $T_{c}=2.11K$ and $T_{c}=13.0K$ for $\mathrm{RuI_{3}}$ and $\mathrm{RuBr_{3}}$, respectively. 
Relatively small $T_{c}$ for the former structure is a direct consequence
of weak $J_{i}$ values of this structure (see S.M Table S3 and S4). $T_{c}$ value obtained for $\mathrm{RuI_{3}}$ monolayer is reasonably below the value reported by Huang 
and coworkers \cite{huang2017quantum}. The reason is straightforward based on two reasons. First,
in Ref. \cite{huang2017quantum}, the authors considered only the nearest neighbor exchange interactions with $J_{1}=82$ meV which is fairly larger than our predicted value. Second, they omitted the effect of MAE
(it seems that MAE is rather influential in $\mathrm{RuX_{3}}$, see Table S.M Table S4) in their calculations. 
On the other hand, $T_{c}$ value obtained for $\mathrm{RuBr_{3}}$ can be compared with $T_{c}=14.21K$ for $\mathrm{RuCl_{3}}$ reported in our recent work \cite{sarikurt2018electronic}. We note that 
recently it has also been  reported for 2D ferromagnetic monolayers $\mathrm{CrX_{3}}$ (X=Br,I) that Curie temperature of $\mathrm{CrBr_{3}}$ is lower than that obtained for $\mathrm{CrI_{3}}$ \cite{liu2016exfoliating}. This is an opposite scenario in comparison to 
our reported values for $\mathrm{RuX_{3}}$ where the Curie temperature of $\mathrm{RuBr_{3}}$ is larger than that of $\mathrm{RuI_{3}}$. These results show that the presence of Ru instead of Cr in monolayer trihalides $\mathrm{MX_{3}(X=Br,I)}$
may cause dramatic differences in critical behavior of these structures. Moreover, as we mentioned before, based on our rigorous DFT calculations, 
we believe that the magnetic behavior of such systems cannot be modeled using standard  Ising model, since the MAE plays a significant role in the magnetic behavior of these materials. Therefore we suggest to use the anisotropic Heisenberg model in atomistic spin model calculations.

Apart from these observations, using the Hamiltonian parameters provided in S.M Table S3 and S4, we have also performed MC simulations for the bulk $\mathrm{RuX_{3}}(\mathrm{X=Br,I})$. 
By assuming weak van der Waals bonding between adjacent magnetic interlayers \cite{sears,banerjee2016proximate,plumb2014alpha}, we followed the same simulation procedure defined for our monolayer systems.
According to our simulation data, we found that the transition
temperatures for $\mathrm{RuI_{3}}$ and $\mathrm{RuBr_{3}}$ in bulk form are given as $T_{c}=0.11K$ and $T_{c}=13.3K$, respectively. 
We note that although the Curie temperature of monolayer $\mathrm{RuBr_{3}}$ is comparable to its bulk counterpart,
the situation is different for $\mathrm{RuI_{3}}$ where the critical temperature of the bulk system is lower than that of the monolayer system. 
This is primarily due to the fact that 
while the values of the exchange interactions for monolayer and bulk cases are in the same order for $\mathrm{RuBr_{3}}$,
the bulk exchange coupling parameters of $\mathrm{RuI_{3}}$ predicted by our DFT calculations have been found to be fairly weaker than those calculated for the monolayer case (c.f. S.M Table S3). 
This means that a small amount of thermal fluctuation can be enough to destroy the magnetic order for the bulk $\mathrm{RuI_{3}}$ system.
Based on a recent experimental work \cite{huang2017layer}, bulk to monolayer transition in $\mathrm{CrI_{3}}$
have been reported with respective transition temperatures $T_{c}=61K$ (bulk) and $T_{c}=45K$ (monolayer). From this point of view, we have an opposite scenario where
our $\mathrm{RuI_{3}}$ system in bulk form exhibits lower critical temperature than that obtained for the monolayer limit. Hence, we can conclude that due to the presence of Ru instead of Cr
in trihalides of the form $\mathrm{MX_{3}(X=Br,I)}$, the bulk magnetic properties may also be significantly altered. This can be a direct consequence of  
different spin magnitudes of Ru and Cr, different exchange energies in the intralayer, as well as interlayer regions, etc.. In conclusion, one cannot establish a general trend for the critical behavior 
(i.e. variation of the critical temperature with the spatial dimension) of $\mathrm{RuX_{3}(X=Br,I)}$ 
when the topology evolves from bulk to monolayer.

\begin{figure*}[!htbp]
	\includegraphics[scale=0.32]{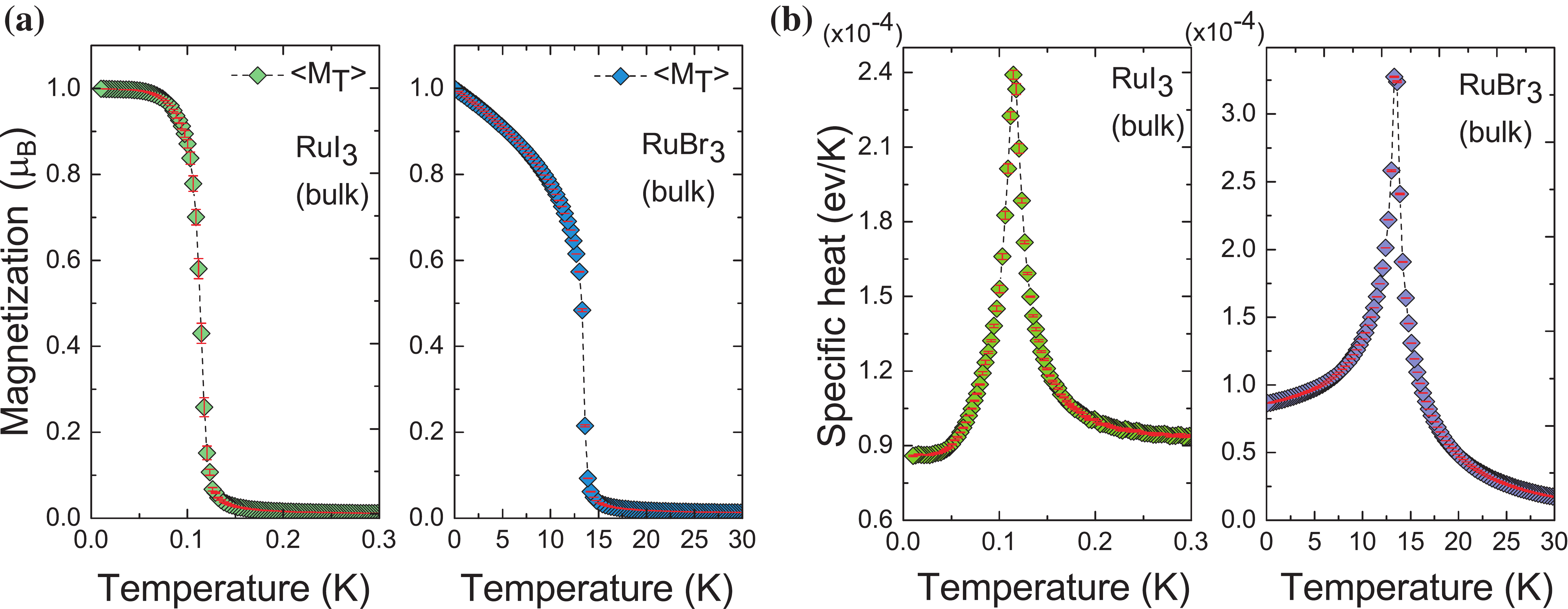}
	\caption{Temperature dependence of (a) average magnetization $M_{T}$ and (b) specific heat for $\mathrm{RuI_{3}}$ and $\mathrm{RuBr_{3}}$ in bulk form. Specific heat curves have been obtained using Eq. (\ref{mc_eq6}).} 
	\label{mc_fig2} 
\end{figure*}

\section{Supplementary Material}
See supplementary material for the exchange interaction parameters such as $J_1$, $J_2$ and $J_3$, phonon band structures of bulk RuX$_3$, electronic 
density of states of monolayer RuX$_3$, relative energy differences for each configurations with respect to U and U+SOC parameters and compared 
band structures of monolayer RuX$_3$ in NM, FM, Neel, Stripy and Zigzag magnetic order using U+SOC methods.

\section{Conclusion}
In conclusion, with the help of first principles calculations we theoretically showed that bulk RuBr$_3$ and RuI$_3$ could be stable in P3$_1$12 space group 
similar to $\alpha$-RuCl$_3$. According to cleavage energy calculations, monolayer forms of RuX$_3$ structures can be easily attained from their bulk phases. 
Also we tested dynamical and thermal stabilities of monolayers and found that they can be stable at room temperature and above. While ferromagnetic spin 
orientation is favorable state for PBE and PBE+SOC calculations, Hubbard U and U+SOC calculations show that AFM-zigzag cases have minimum ground state energies 
comparing to others except for U=0.5 eV. However, electronic band structures of all spin oriented configurations show similarities, U and SOC effects 
enhance the band gaps. While RuI$_3$ monolayer has band gaps values in the range of infrared region, band gap values of RuBr$_3$ monolayer can reach the near visible region according to spin orientation configuration and U parameter. 
We have also performed detailed Monte Carlo simulations to clarify the magnetic properties of $\mathrm{RuBr_{3}}$ and $\mathrm{RuI_{3}}$. Using the atomistic model parameters (i.e. exchange and magnetic anisotropy energies) obtained from PBE+SOC calculations, we have found that the Curie temperature of $\mathrm{RuBr_{3}}$ dominates against that of $\mathrm{RuI_{3}}$ both in bulk and monolayer forms. However, obtained critical temperature values are found to be far from the room temperature. Furthermore, some drastic changes may originate in the magnetic behavior of these systems when the form is changed from bulk to monolayer. According to the DFT calculations based on U+SOC, ground state configuration evolves from ferromagnetic to antiferromagnetic zigzag which causes prominent changes in the numerical values of simulation parameters (c.f. see S.M Tables S3, S6, S7). Besides, since the magnetic character of the first, second and third nearest neighbors turn into AFM type, RuX$_3$ system in bulk and monolayer phases exhibits Neel temperature instead of Curie temperature. More importantly, frustration effects take place in the system which completely affects the magnetic behavior. We should also note that, by comparing the magnetic anisotropy constants in the presence of U+SOC, we see that in the monolayer case, absolute values of the anisotropy constants attain lower values, in comparison to the case of PBE+SOC. In addition, in the monolayer case, anisotropy constants also lose their in plane isotropy (c.f. compare the numerical values of $k_x$ and $k_y$ between S.M. Tables S4 and S5). Overall, the entire magnetic behavior of RuX$_3$ (X=Br, I) may be highly sensitive to the consideration of Hubbard U parameter in DFT calculations.
We believe that this study can play an important role, for the future attempt to obtain bulk and monolayer forms of RuBr$_3$ and RuI$_3$.


\begin{acknowledgments}
This work was supported by the Scientific and Technological Research Council of Turkey (TUBITAK) under the Research Project No. 117F133. Computing resources used in this work were provided by the TUBITAK ULAKBIM, High Performance and Grid Computing Center (Tr-Grid e-Infrastructure) and the National Center for High Performance Computing of Turkey (UHeM) under grant number 5004972017.
\end{acknowledgments}




\newpage
\providecommand{\noopsort}[1]{}\providecommand{\singleletter}[1]{#1}%

\newpage
\section{Supplementary Material:Exploring the Electronic and Magnetic Properties of New Metal Halides from Bulk to Two-Dimensional Monolayer: RuX$_3$ (X=Br, I)}
\subsection{Bulk RuX$_3$}
\begin{figure*}[!h]
	\includegraphics[scale=0.8]{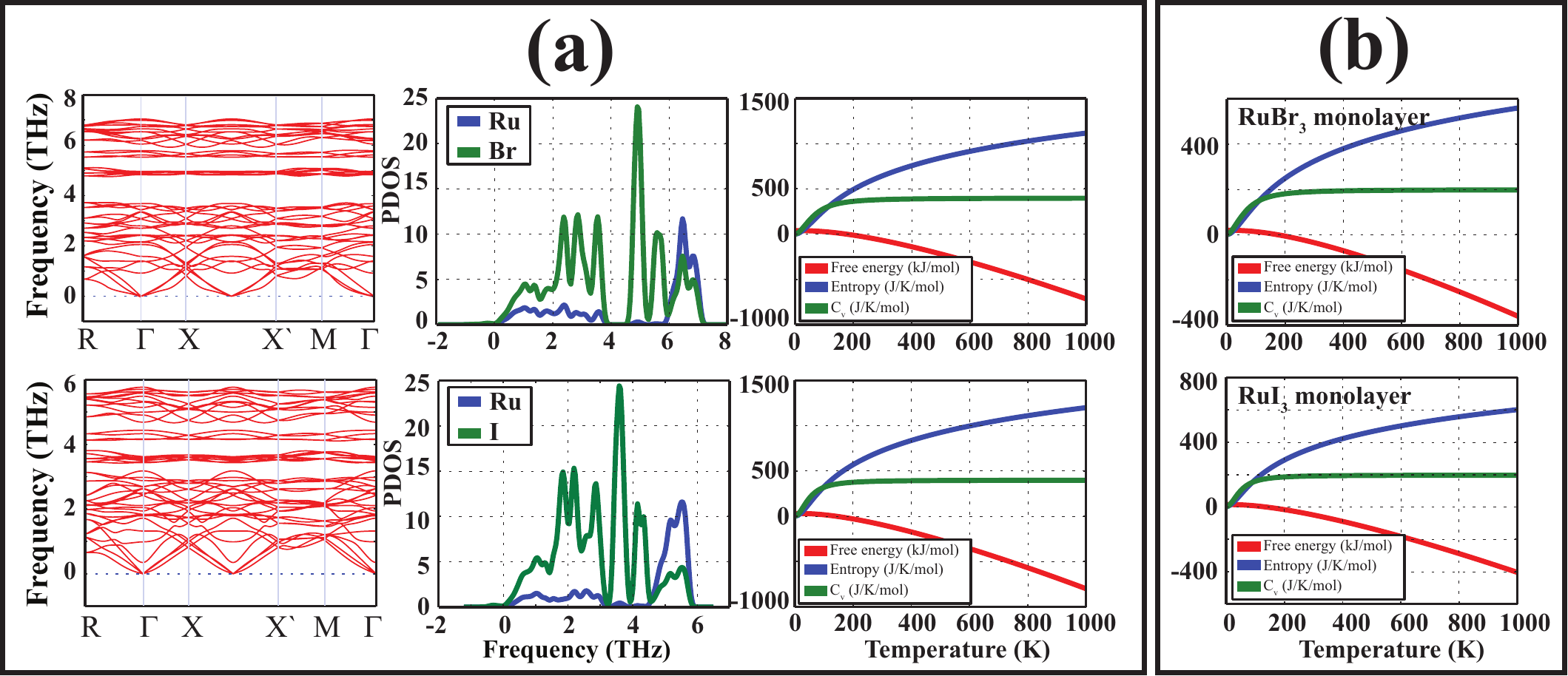}
	\caption{(Color online) a) Phonon band structure, phonon partial density of states and thermodynamic properties as 
		a function of temperature of bulk (P3$_1$12 space group) RuBr$_3$ and RuI$_3$ structures, b) thermodynamic properties as a function of temperature of 
		monolayer RuBr$_3$ and RuI$_3$ structures.}
	\label{fig_bandpdos}
\end{figure*}

Figure~\ref{fig_bandpdos} illustrates the phonon band structure of bulk RuX$_3$ structures with comparable density of states and thermodynamic variables 
of their bulk and 
monolayer phases. Phonon dispersions show that all phonon modes have positive value in the whole Brillouin zone, which imply the dynamical 
stability at T$\sim$0K. As can be seen from phonon DOS, there are only Br atoms vibrations in the range of 5-6 THz for RuBr$_3$ and I atoms vibrations 
in the range of 3-4.5 THz for RuI$_3$ bulk structures. Probably these dominant peaks could be seen in Raman spectrums. Thermodynamic variables are extracted 
from PHONOPY after the phonon calculations and given as a function of temperature. As can be seen, free energies of bulk and monolayer of RuX$_3$ structures 
go to negative values after about 200K, and also C$_v$, heat capacities become fixed for T$>$200K and tend to the Dulong-Petit limit.  

\begin{figure*}[!h]
	\includegraphics[scale=0.4]{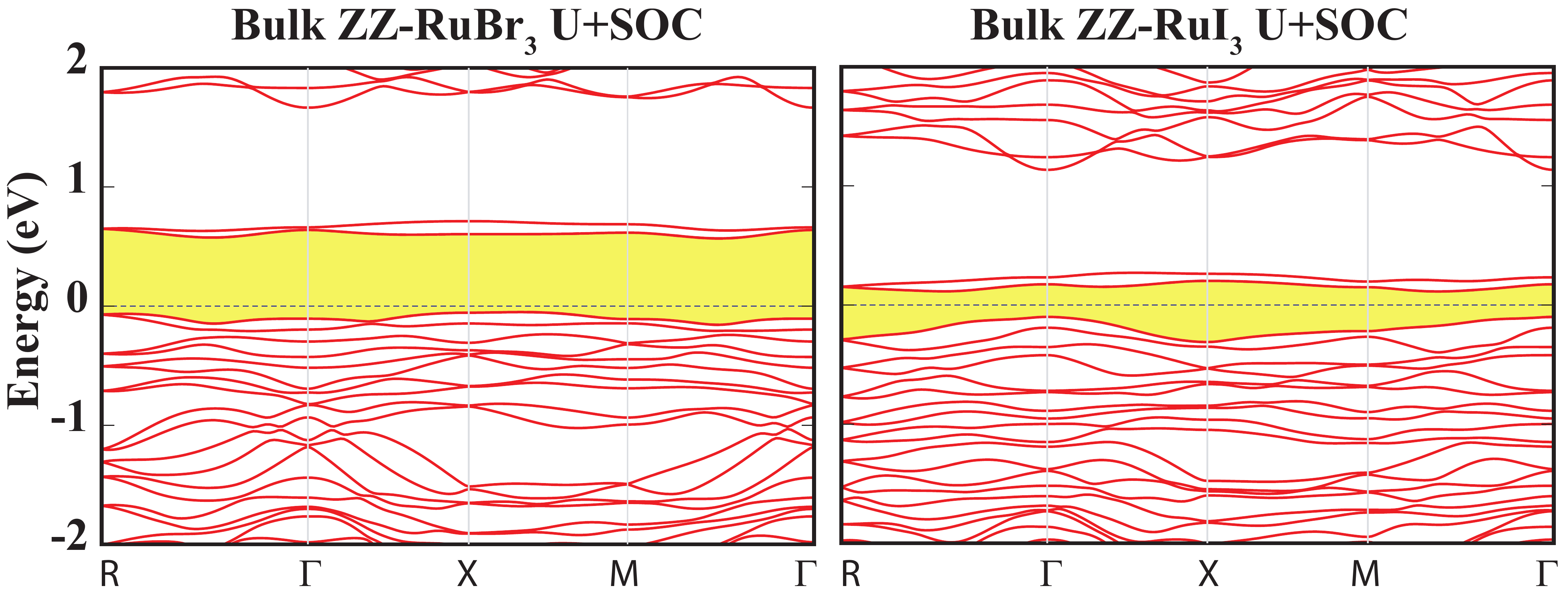}
	\caption{(Color online) Electronic band structure of bulk RuX$_3$ (X=Br,I) in Zigzag magnetic order using U+SOC methods (U=1.5 eV). 
		Fermi energy is ste to zero energy and band gaps are colored.}
	\label{fig_bulkband}
\end{figure*}

We determine the favorable magnetic ground state of RuX$_3$ (X=Br,I) bulk structures with considering the Hubbard U+SOC (U=1.5 eV) effect. As in the case for RuX$_3$ monolayer, we reveal that Zigzag spin orientation is energetically stable for bulk structures.
We also performed electronic band structure calculation for RuX$_3$ (X=Br,I) bulk structures with Hubbard U+SOC (U=1.5 eV) effect (Figure~\ref{fig_bulkband}). We find out the energy band gaps as 0.62 eV and 0.22 eV for RuBr$_3$ and RuI$_3$, respectively.

\subsection{Spin orientation and exchange interaction energy}
\begin{figure*}[!h]
	\includegraphics[scale=0.35]{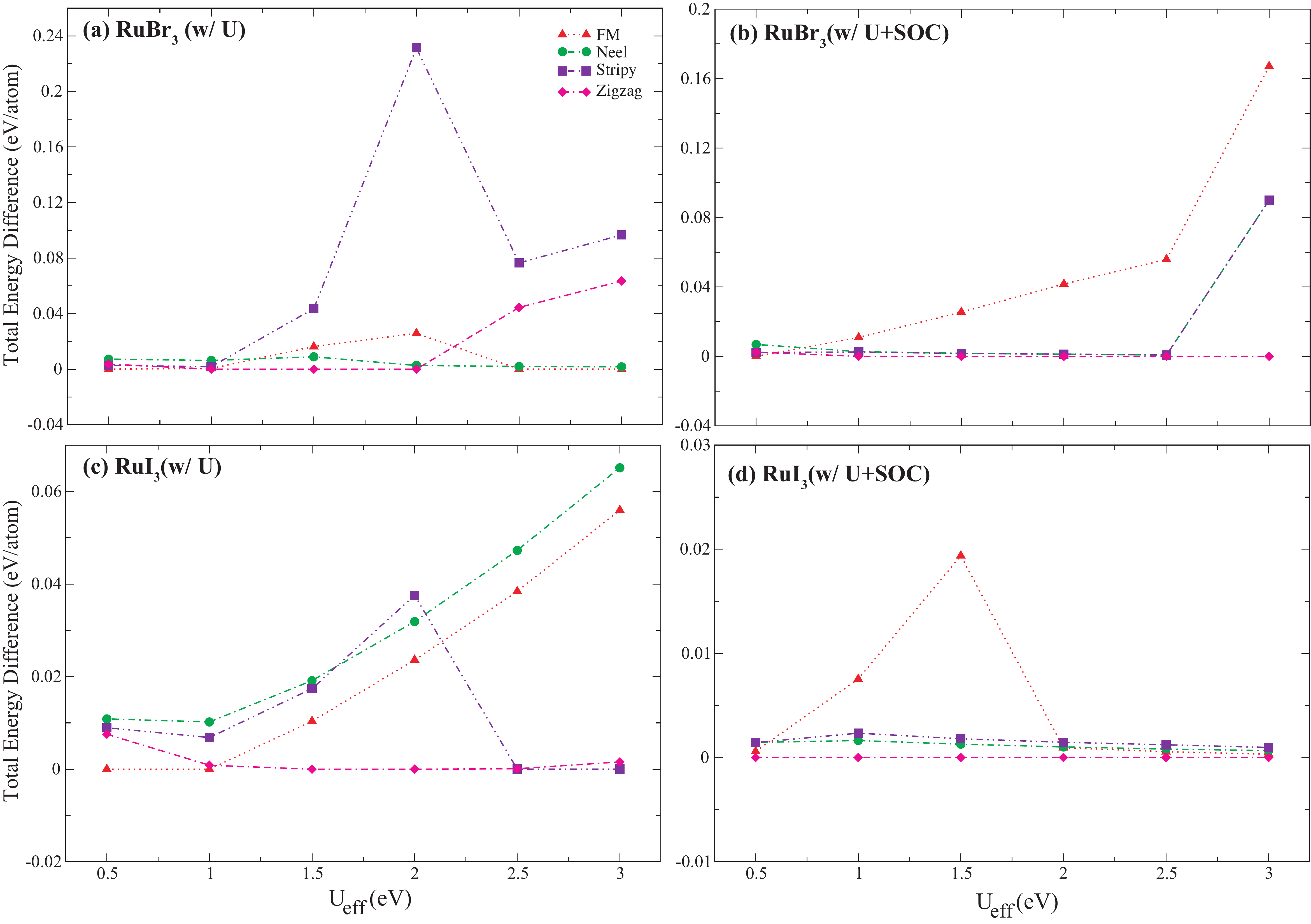}
	\caption{(Color online)Relative energy difference for each configuration with respect to U$_{eff}$ using the U and U+SOC methods. (a)-(b) RuBr$_3$ monolayer (c)-(d) RuI$_3$ monolayer. Zero of energy refers the ground state energy.}
	\label{fig_relative}
\end{figure*}

Since free Ruthenium and each halogen atoms has magnetization (2$\mu_B$ for Ru, and 1 $\mu_B$ for halogens), spin orientation in the structure which includes 
these atoms, will have an important role to determine the ground state energy of the system. Hence, we optimized all structures for various Hubbard U 
parameters with and without SOC parameters for all considered spin orientations and obtained their ground state energies. Figure~\ref{fig_relative} illustrates the relative 
ground state energies with respect to Hubbard parameters. Generally for small U parameter (U=0.5 eV) ferromagnetic spin orientation is favorable state, but 
with increasing U value zigzag orientation becomes energetically favorable. We also calculated exchange interaction energy for PBE, PBE+SOC, U and U+SOC 
calculations as seen in Table~\ref{table3_bulk},~\ref{tableBr3_exc_usoc},~\ref{tableI3_exc_usoc} . These extended calculations show that exchange interaction 
energy increases with increasing of U energy. 

\subsection{Electronic properties of RuX$_3$ monolayers}
\begin{figure*}[!h]
	\includegraphics[scale=0.35]{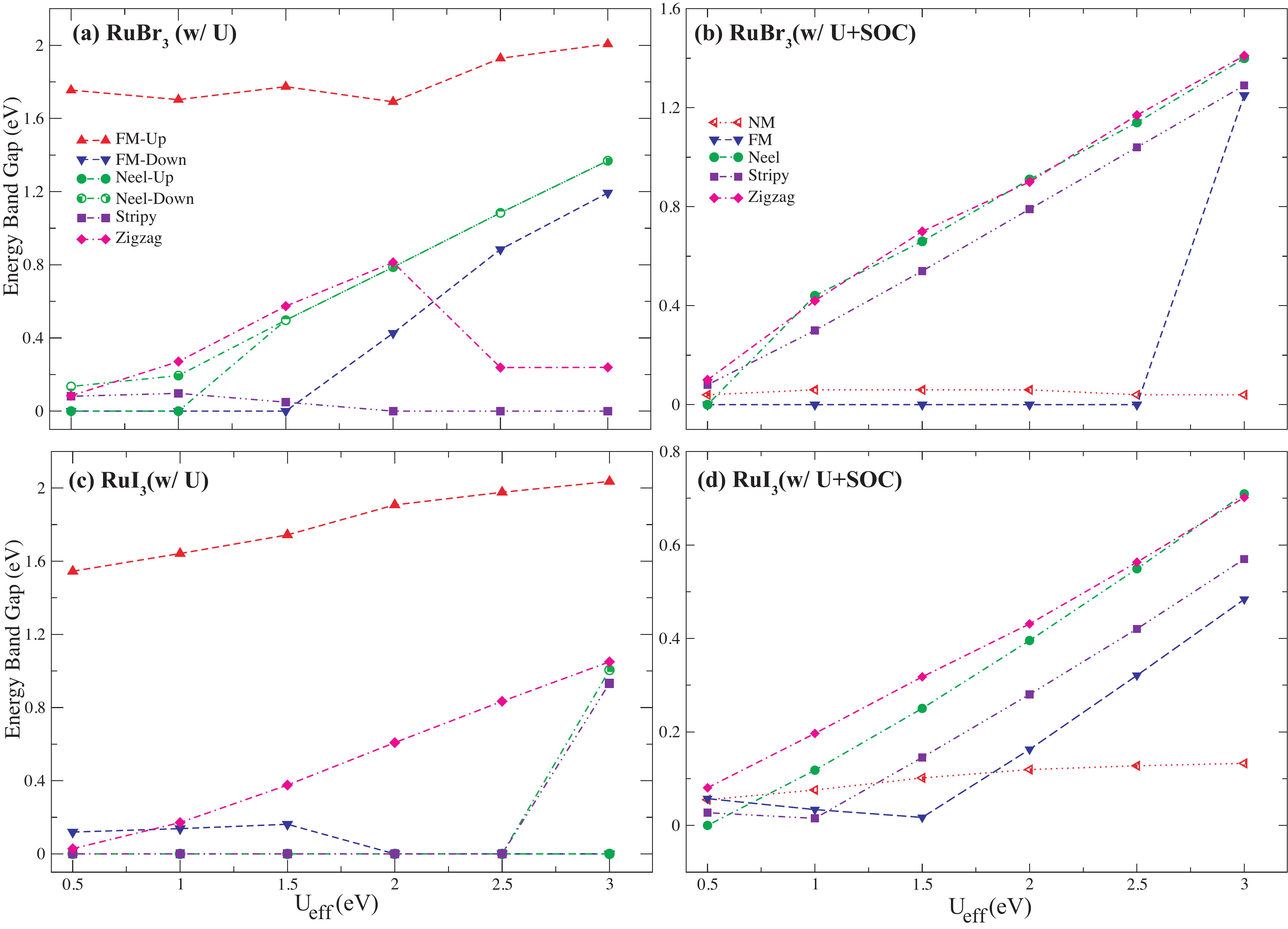}
	\caption{(Color online) Calculated energy band gap values of all magnetic configurations as a function of U$_{eff}$ using the U and U+SOC methods. U results are depicted in (a) and (c) for RuBr$_3$ and RuI$_3$, respectively. U+SOC results are given in (b) and (d) for RuBr$_3$ and RuI$_3$, respectively.}
	\label{fig_egap}
\end{figure*}

\begin{figure*}[!h]
	\includegraphics[scale=0.15]{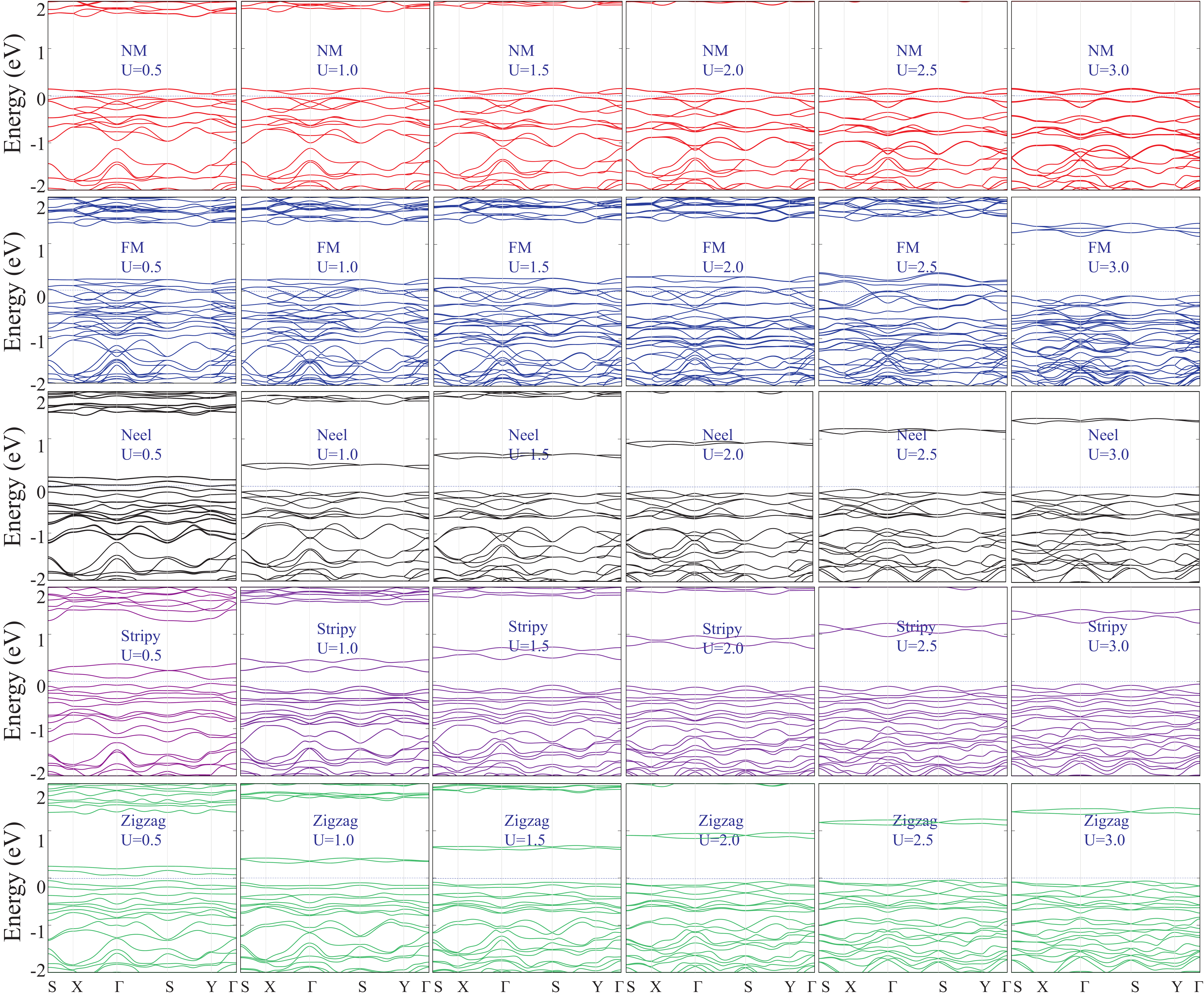}
	\caption{(Color online) The effect of on-site Coulomb interaction on the electronic structure of monolayer RuBr$_3$ in NM, FM (only for spin-up channels), Neel, Stripy and Zigzag magnetic order using U+SOC methods.}
	\label{fig_RuBr3egap}
\end{figure*}

\begin{figure*}[!h]
	\includegraphics[scale=0.15]{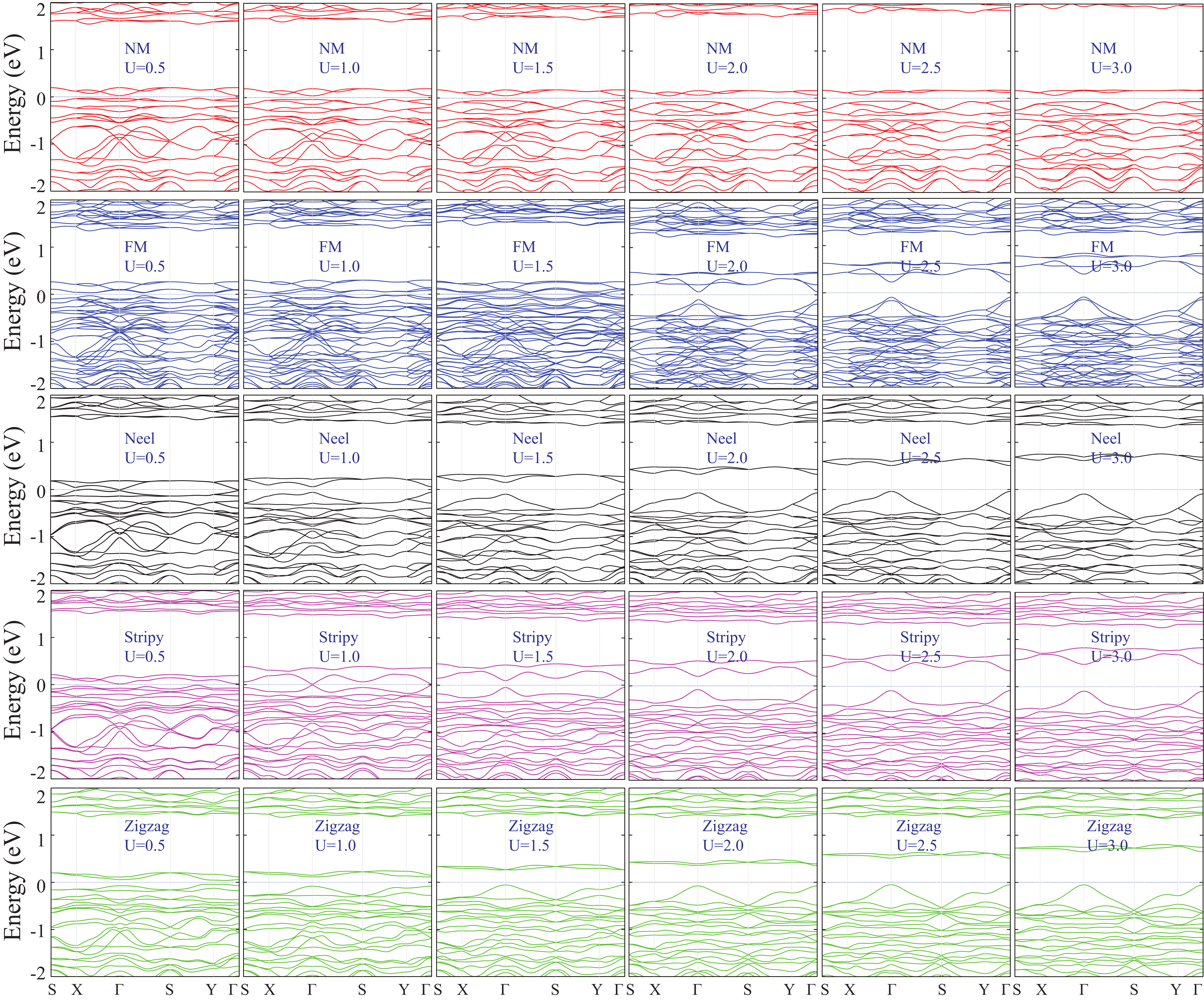}
	\caption{(Color online) The effect of on-site Coulomb interaction on the electronic structure of monolayer RuI$_3$ in NM, FM (only for spin-up channels), Neel, Stripy and Zigzag magnetic order using U+SOC methods.}
	\label{fig_RuI3egap}
\end{figure*}

Figure~\ref{fig_egap},~\ref{fig_RuI3egap},~\ref{fig_RuBr3egap} show the band gap trend and band structures of all considered spin oriented RuX$_3$ 
monolayers for various U and U+SOC parameters. As can be seen from the band structures, increasing of U value enhances the band gaps. For FM spin oriented 
structures Hubbard U Coulomb interaction effective after U=2 eV for RuBr$_3$, while this U value is 3 eV for RuI$_3$ monolayer. FM band structures are plotted for 
only spin-up channels due to spin-down channels have huge band gaps values, which are very out of the scales that plotted.

\begin{table*}[!h]
	\caption{Calculated relative energy values in meV/atom (E$_{rel}$) which are obtained from the PBE, PBE+SOC and U+SOC (U=1.5 eV) calculations for bulk RuX$_3$ (X=Br,I). Zero of energy refers the ground state energy.}
	\label{table1}
	\renewcommand{\arraystretch}{1.5}
	\begin{tabular}{cc|c|c|c|c}
		\hline
		\hline
		&   &  FM  & Neel & Stripy & Zigzag \\
		\hline 
		RuBr$_{3}$  & E$_{rel} ^{PBE}$ &  0 &  $13.63$\qquad & $5.58$\qquad & $6.69$  \\
		\hline
		& E$_{rel} ^{PBE+SOC}$ &  0  & $9.46$\qquad &$4.86$ \qquad & $4.22$  \\
		\hline 
		& E$_{rel} ^{U+SOC}$ &  $20.82$\qquad  & $2.00$\qquad & $1.89$\qquad & 0  \\
		\hline 
		RuI$_{3}$  & E$_{rel} ^{PBE}$ & 0 & $1.99$\qquad & $1.86$\qquad & $1.76$  \\
		\hline
		&  E$_{rel} ^{PBE+SOC}$ &  0 & $0.02$\qquad &$0.02$ \qquad & $0.02$  \\
		& E$_{rel} ^{U+SOC}$ & $9.5$\qquad & $0.18$\qquad &$8.69$\qquad & 0 \\
		\hline 
		\hline
		\hline
	\end{tabular}
\end{table*}

\begin{table*}[!h]
	\caption{Calculated relative energy values in meV/atom (E$_{rel}$) which are obtained from the PBE and PBE+SOC calculations for monolayer
		RuX$_3$ (X=Br,I). Zero of energy refers the ground state energy.}
	\label{table2}
	\renewcommand{\arraystretch}{1.5}
	\begin{tabular}{cc|c|c|c|c|c}
		\hline
		\hline
		&   &  FM &  NM & Neel & Stripy & Zigzag \\
		\hline 
		RuBr$_{3}$  & E$_{rel}^{PBE}$ &  0 &  $22.67$\qquad & $21.58$\qquad & $5.38$\qquad & $7.52$  \\
		\hline
		& E$_{rel} ^{PBE+SOC}$ &  0 & $14.68$\qquad & $13.07$\qquad &$4.19$ \qquad & $4.80$  \\
		\hline 
		RuI$_{3}$  & E$_{rel} ^{PBE}$ &  0 &  $10.08$\qquad & $10.07$\qquad & $7.45$\qquad & $9.48$  \\
		\hline
		&  E$_{rel} ^{PBE+SOC}$ &  0 & $0.66$\qquad & $0.66$\qquad &$0.65$ \qquad & $0.65$  \\
		\hline
		\hline
	\end{tabular}
\end{table*}

\begin{table*}[!h]
	\caption{Calculated exchange interaction parameters $J_1$, $J_2$, $J_3$ (meV) for bulk and monolayer RuX$_3$ (X=Br,I) from the PBE and PBE+SOC calculations.}
	\label{table3_bulk}
	\renewcommand{\arraystretch}{1.5}
	\begin{tabular}{cc|c|c|c}
		\hline  
		\hline
		&  & $J_1$  & $J_2$ & $J_3$ \\
		\hline
		RuBr$_{3}$ (Bulk) & PBE & $6.26$ &  $-0.34$ &  $2.82$ \\
		\hline
		& PBE+SOC & $5.05$ & $-0.093$ &  $1.25$\\
		\hline
		& U+SOC (U=1.5 eV) & $-8.46$ & $-5.23$ &  $-4.08$\\
		\hline
		RuBr$_{3}$ (ML) & PBE & $9.72$ &  $-2.17$ &  $4.66$ \\
		\hline
		& PBE+SOC & $6.23$ & $-1.02$ &  $2.48$\\
		\hline
		RuI$_{3}$ (Bulk) & PBE & $1.11$ &  $0.44$ &  $0.21$ \\
		\hline
		& PBE+SOC & $11.1\times 10^{-3}$ & $4.02\times 10^{-3}$ &  $1.64\times 10^{-3}$\\
		\hline
		& U+SOC (U=1.5 eV) & $-0.31$ & $-0.25$ &  $-5.9$\\
		\hline
		RuI$_{3}$ (ML) & PBE & $4.02$ &  $1.72$ &  $2.69$ \\
		\hline
		& PBE+SOC & $0.33$ & $0.16$ &  $0.11$\\
		\hline
		\hline
	\end{tabular}
\end{table*}

\begin{table*}[!h]
	\caption{Magneto-crystalline anisotropy energies and anisotropy constants for FM configuration of bulk and monolayer RuX$_3$ (X=Br,I) using PBE+SOC method.}
	\label{table5_bulkmae}
	\renewcommand{\arraystretch}{1.5}
	\begin{tabular}{l|c|c|c|c}
		\hline 
		\hline 
		&  out-of-plane &   in-plane   & \quad $k_{x}$ \quad & \quad $k_{y}$ \quad \\
		& $E[100]-E[001]$ &   $E[100]-E[010]$  &  (eV) & (eV)  \\
		\hline 
		\hline 
		RuBr$_3$ (Bulk) &  $-7.26\ meV$ &  $-1.5\ meV$ & $-0.0073$ & $-0.0088$ \\
		\hline 
		RuBr$_3$ (ML) &  $-16.65\ meV$ &  $-16\ \mu eV$ & $-0.017$ & $-0.017$ \\
		\hline 
		RuI$_3$ (Bulk) &  $-17.5\ meV$ &  $1.45\ meV$ & $-0.0175$  & $-0.019$\\
		\hline 
		RuI$_3$ (ML) &  $-29.04\ meV$ &  $58\ \mu eV$ & $-0.029$  & $-0.029$ \\
		\hline
		\hline 
	\end{tabular}
\end{table*}

\begin{table*}[!h]
	\caption{Magneto-crystalline anisotropy energies and anisotropy constants for Zigzag configuration of bulk and monolayer RuX$_3$ (X=Br,I) using U+SOC method (U=1.5 eV).}
	\label{table6_bulkmae}
	\renewcommand{\arraystretch}{1.5}
	\begin{tabular}{l|c|c|c|c}
		\hline 
		\hline 
		&  out-of-plane &   in-plane   & \quad $k_{x}$ \quad & \quad $k_{y}$ \quad \\
		& $E[100]-E[001]$ &   $E[100]-E[010]$  &  (eV) & (eV)  \\
		\hline 
		\hline 
		RuBr$_3$ (Bulk) &  $-5.26\ meV$ &  $-0.36\ meV$ & $-0.0053$ & $-0.0049$ \\
		\hline 
		RuBr$_3$ (ML) &  $-5.26\ meV$ &  $-0.35\ meV$ & $-0.0052$ & $-0.0049$ \\
		\hline 
		RuI$_3$ (Bulk) &  $-13.19\ meV$ &  $\ 1.21 meV$ & $-0.0144$  & $-0.0132$ \\
		\hline 
		RuI$_3$ (ML) &  $-12.88\ meV$ &  $-4.27\ meV$ & $-0.0129$  & $-0.0086$ \\
		\hline
		\hline 
	\end{tabular}
\end{table*}

\begin{table*}[!h]
	\caption{Calculated exchange interaction parameters $J_1$, $J_2$, $J_3$ (meV) for monolayer RuBr$_{3}$ using the U and U+SOC methods.}
	\label{tableBr3_exc_usoc}
	\renewcommand{\arraystretch}{1.5}
	\begin{tabular}{c|c|c|c||c|c|c|}
		\hline  
		\hline
		&  \multicolumn{3}{c||}{ w/o SOC}  & \multicolumn{3}{c|}{ w SOC} \\
		\hline
		U & $J_1$ &  $J_2$ & $J_3$ &  $J_1$ &  $J_2$ & $J_3$  \\
		\hline
		0.5 & 3.18 & -0.19 & 1.64 & 4.46 & -1.11 & 1.62  \\
		\hline
		1.0 & 3.79 & -1.25 & 0.07 & -2.82  & -2.77 & -2.63 \\
		\hline
		\textbf{1.5} & \textbf{18.15} & \textbf{4.63} & \textbf{-23.05} & \textbf{-11.04}  & \textbf{-6.40} & \textbf{-4.78} \\
		\hline
		2.0 & 17.34 & 7.34 & -32.79 & -19.56  & -10.41 & -7.39 \\
		\hline
		2.5 & 17.03 & 29.76 & -15.73 & -27.20  & -13.96  & -9.57 \\
		\hline
		3.0 & 17.42 & 39.63 & -16.29 & -6.40 & -41.78 & -57.82 \\
		\hline
		\hline
	\end{tabular}
\end{table*}

\begin{table*}[!h]
	\caption{Calculated exchange interaction parameters $J_1$, $J_2$, $J_3$ (meV) for monolayer RuI$_{3}$ using the U and U+SOC methods.}
	\label{tableI3_exc_usoc}
	\renewcommand{\arraystretch}{1.5}
	\begin{tabular}{c|c|c|c||c|c|c|}
		\hline  
		\hline
		&  \multicolumn{3}{c||}{ w/o SOC}  & \multicolumn{3}{c|}{ w SOC} \\
		\hline
		U & $J_1$ &  $J_2$ & $J_3$ &  $J_1$ &  $J_2$ & $J_3$  \\
		\hline
		0.5 & 6.13 & 1.42 & 1.10 & 1.15 & -0.15 & -0.58  \\
		\hline
		1.0 & 8.08 & -0.63 & -1.28 & -1.78 & -1.71 & -2.15 \\
		\hline
		\textbf{1.5} & \textbf{13.09} & \textbf{-3.02} & \textbf{-7.24}  & \textbf{-8.14} & \textbf{-4.71} & \textbf{-3.92} \\
		\hline
		2.0 & 22.94 & -4.48 & -17.41 & 0.77 & -0.12 & -0.72 \\
		\hline
		2.5 & 4.379 & -21.39 & 1.52 & 0.74 & -0.038 & -0.57 \\
		\hline
		3.0 & 3.80 & -29.88  & 2.31 & 0.66 & -0.004 & -0.42 \\
		\hline
		\hline
	\end{tabular}
\end{table*}


\begin{figure*}[!h]
	\includegraphics[scale=0.8]{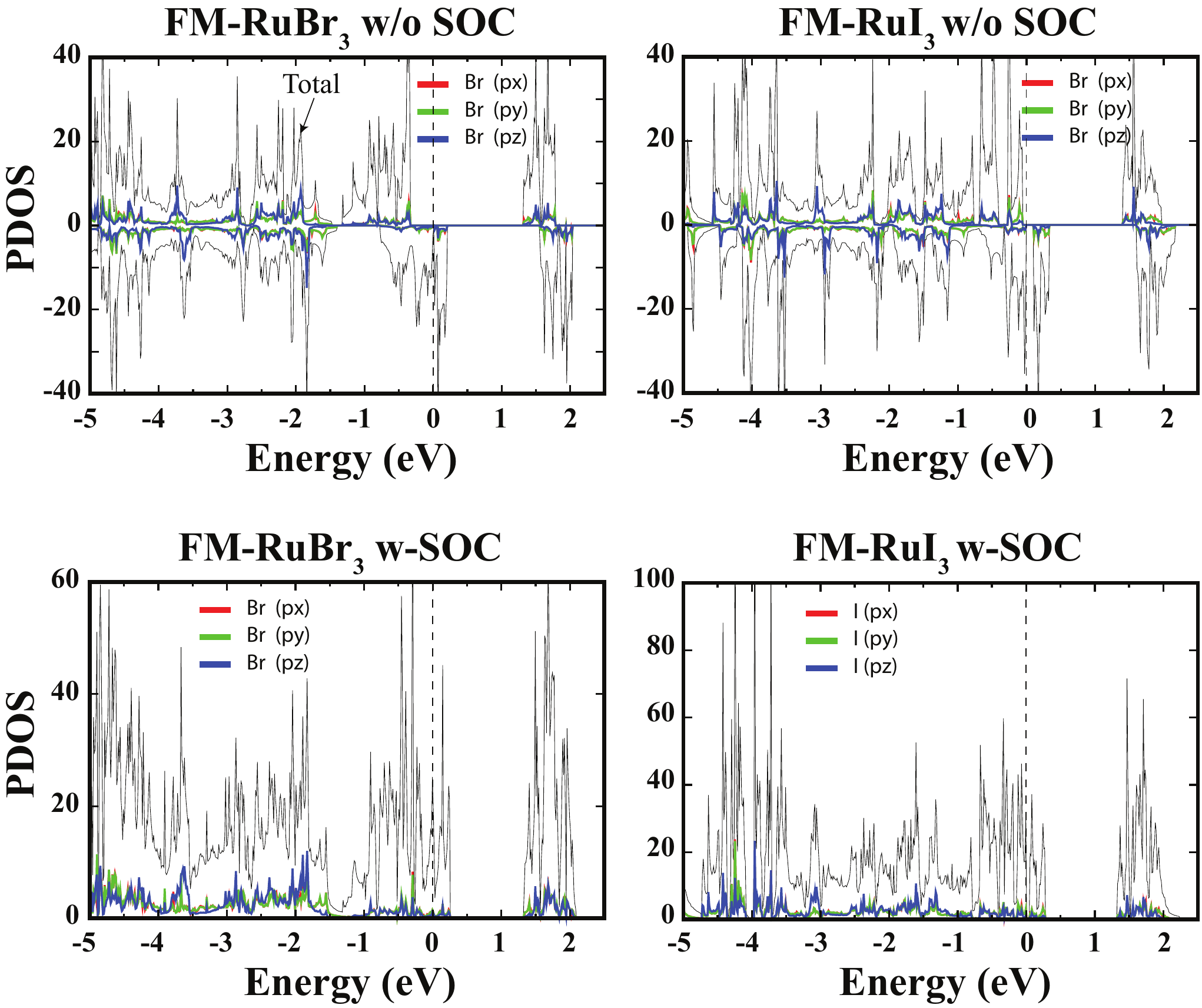}
	\caption{(Color online) Electronic partial density of states of RuBr$_3$ and RuI$_3$ monolayers, which are obtained from the PBE and PBE+SOC calculations.}
	\label{fig_pdos}
\end{figure*}

\end{document}